\documentclass[twocolumn,aps,amsmath,amssymb,showpacs,floatfix]{revtex4}
\usepackage{graphicx}
\usepackage{dcolumn}
\usepackage{bm}
\usepackage{float}
\begin{document}
\title{Dynamics of an Unbounded Interface Between Ordered Phases}
\author{P.~L.~Krapivsky}
\email{paulk@bu.edu}
\author{S.~Redner}
\email{redner@bu.edu}
\affiliation{Center for BioDynamics, Center for Polymer Studies, 
and Department of Physics, Boston University, Boston, MA 02215, USA}
\author{J. Tailleur}
\email {Julien.Tailleur@crans.org}
\affiliation{Departement de Physique, ENS de Cachan, 
61 avenue du President Wilson, 94235 Cachan Cedex, Paris, France}
\begin{abstract} 
  We investigate the evolution of a single unbounded interface between
  ordered phases in two-dimensional Ising ferromagnets that are endowed with
  single-spin-flip zero-temperature Glauber dynamics.  We examine
  specifically the cases where the interface initially has either one or two
  corners.  In both examples, the interface evolves to a limiting
  self-similar form.  We apply the continuum time-dependent Ginzburg-Landau
  equation and a microscopic approach to calculate the interface shape.  For
  the single corner system, we also discuss a correspondence between the
  interface and the Young diagram that represents the partition of the
  integers.
\end{abstract}
\maketitle 

\section{introduction} 

At low temperatures, interfaces between two broken-symmetry ordered phases
typically shrink and eventually disappear \cite{rev}.  The dynamics is
usually driven by forces that reduce the interface and leads to surprisingly
complicated coarsening processes --- even the evolution of an isolated
simply-connected domain of minority phase in a sea of the majority phase is
in general insoluble.  However, every finite domain of linear size $R$
disappears in a finite time that scales as $R^2$ \cite{lif} for dynamics that
does not conserve the order parameter \cite{glauber}.  In this sense, we
understand the shrinking of a single domain, or equivalently, the evolution
of a single bounded interface \cite{note}.

\begin{figure}[ht] 
 \vspace*{0.cm}
 \includegraphics*[width=0.48\textwidth]{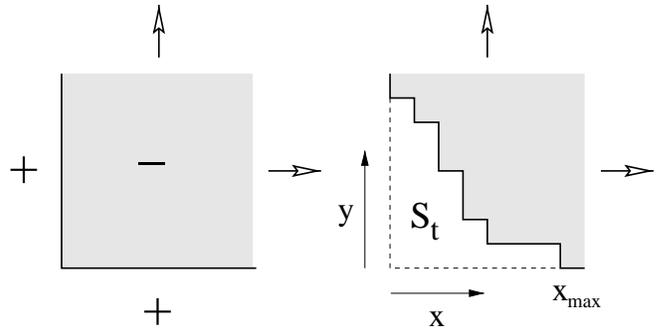}
\caption{A single corner interface (left) in the initial state and some
  time later (right).  Grey denotes spin down (extending to $\infty$ in
  the $+x$ and $+y$ directions), and white denotes spin up.  The evolving
  interface encloses an area $S_t$ at time $t$. 
\label{def}}
\end{figure}

The goal of this work is to understand the evolution of a single {\em
  unbounded\/} two-dimensional interface in simple geometric configurations.
The most elementary such example is an infinite straight interface.  For this
geometry, any spin flip event increases the length of the interface and
raises the energy.  Thus a straight interface does not evolve --- to have any
evolution at zero temperature, the interface must have curvature.
The simplest realization of a curvature in a lattice system is an infinite
interface with a single corner (Fig.~\ref{def}).  While not a direct analog
of the theoretical models we consider in this paper, a physical realization
of such a geometry is the spreading of a fluid in a V-shaped groove \cite{W}.

According to zero-temperature single-spin-flip Glauber dynamics
\cite{glauber}, the corner spin $\sigma_{0,0}$ can flip.  After it flips, the
neighbors $\sigma_{0,1}$ and $\sigma_{1,0}$ can flip, or $\sigma_{0,0}$ can
flip back, {\it etc}.  The interface thus evolves stochastically and can, in
principle, return to its original configuration as time increases
\cite{convex}.  This is, however, exceedingly improbable and generically the
interface recedes diffusively (since the dynamics does not conserve the order
parameter), {\it i.e.}, $x\propto \sqrt{t}$ and $y\propto \sqrt{t}$.
Furthermore, although the interface at a fixed time fluctuates from
realization to realization, it becomes progressively less random as time
increases.  More precisely, after the contraction $(x,y)\to
(x/\sqrt{t},y/\sqrt{t})$ the interface approaches a deterministic limiting
shape.  In the following two sections, we will study the time evolution of
the interface and determine its shape within a continuum (Sec.~II) and a
microscopic approach (Sec.~III).  In addition to the wedge geometry, we will
study, in Sec.~IV, interface evolution in systems that initially contain two
corners --- the macroscopic step and the semi-infinite finger geometries.
Finally, in Sec.~V, we study the time dependence of fluctuations in the
interface shape.

\section{Coarse-Grained Description}
\label{landau} 

A natural way to study interface evolution is through the time-dependent
Ginzburg-Landau (TDGL) equation \cite{rev}.  It is generally believed that
the long-time behavior predicted by the continuum TDGL equation should be the
same as that for the microscopic Ising-Glauber model.  In the case of the
single interface, we find, surprisingly, that the predictions of the TDGL
description qualitatively disagree with simulations of the Ising interface
that evolves by zero-temperature Glauber kinetics.

Since we are primarily interested in interfacial behavior, we will not write
the TDGL equation but instead will merely utilize a reduction directly to the
interface dynamics.  As found by Allen and Cahn \cite{AC}, the normal
velocity of the interface is proportional to the local curvature, that is,
\begin{equation}
\label{cahn}
v_n=-D\nabla\cdot {\bf n},
\end{equation}
where $D$ is the diffusion constant and {\bf n} is the local normal to the
interface \cite{det}.  This interface dynamics in the TDGL equation is a
specific example of curvature-driven flow \cite{Mu,H84,GH86,G87,CS}, where
one is concerned with the evolution of general shapes in arbitrary dimension
due to a local velocity that is proportional to the local curvature.

For the case of interest to us, namely, a one-dimensional interface whose
locus is $y(x,t)$, the curvature is
\begin{equation}
\label{curv}
\nabla\cdot {\bf n}=-\frac{y_{xx}}{\left[1+y_x^2\right]^{3/2}}\,,
\end{equation}
where the subscripts denote partial differentiation.  Using the kinematic
condition $v_n \sqrt{1+y_x^2}=y_t$ we find that the interface $y(x,t)$ obeys
the diffusion-like equation
\begin{equation}
\label{y}
y_t=D\,\frac{y_{xx}}{1+y_x^2}\,.
\end{equation}
Because of the absence of any constant with dimension of length in this
equation, the corresponding solution admits the self-similar form
\begin{equation}
\label{sim}
y(x,t)=\sqrt{Dt}\,\,Y(X), \quad X=x/\sqrt{Dt}.
\end{equation}
Note that the increase of the magnetization is equal to twice the area under the
curve $y(x,t)$.  {}From ansatz (\ref{sim}), the growth of the area is
proportional to $t$, so that the magnetization also grows linearly with time.
 
To solve the equation of motion (\ref{y}), we substitute into this equation
the ansatz of Eq.~(\ref{sim}) and find that the scaling function $Y(X)$ obeys
\begin{equation}                                                             
\label{Y}
\frac{Y-X Y'}{2}=\frac{Y''}{1+(Y')^2},
\end{equation}
where prime indicates differentiation with respect to $X$.  Equation
(\ref{Y}) should be solved subject to the constraints
\begin{equation}                                                             
\label{bound}
\lim_{X\to\infty} Y(X)=0,\qquad 
\lim_{X\to +0} Y(X)=\infty. 
\end{equation}
Thus we recast the original problem in Eq.~(\ref{y}) into an ordinary
differential equation subject to the above boundary conditions.  Within the
TDGL equation framework we note that one can also study the evolution of a
wedge with an arbitrary opening angle.  For example, if the wedge initially
occupies the region $y>|x|\tan\theta$ we should solve Eq.~(\ref{Y}) subject
to the boundary condition $Y\to \pm X \tan\theta$ as $X\to\pm\infty$.

To solve Eq.~(\ref{Y}), we first introduce the polar coordinates
$(X,Y)=(r\cos \theta, r\sin \theta)$ and after straightforward variable
transformations we recast Eq.~(\ref{Y}) into the following equation for
$r=r(\theta)$:
\begin{equation}                                                             
\label{r}
2r\,\frac{d^2 r}{d\theta^2} 
-\left(4+r^2\right)\left(\frac{d r}{d\theta}\right)^2=r^2\left(2+r^2\right).
\end{equation}
Writing $\frac{d r}{d\theta}=R(r)$, further reduces Eq.~(\ref{r}) to the
first-order equation
\begin{equation}                                                             
\label{R}
\left(r\,\frac{d}{d r}-r^2-4\right)R^2=r^2\left(2+r^2\right),
\end{equation}
whose solution is
\begin{equation}                                                             
\label{Rsol}                                                             
R^2=r^4\,e^{r^2/2}\,F(r,r_*), 
\end{equation}
with
\begin{equation}                                                             
\label{F}                                                             
F(r,r_*)=\int_{r_*}^r d\rho\,\left(
\frac{2}{\rho^3}+\frac{1}{\rho}\right)\,e^{-\rho^2/2}\,.
\end{equation}
The interface is now determined from
\begin{equation}                                                             
\label{rtheta}
\frac{d r}{d\theta}=-r^2\,e^{r^2/4}\sqrt{F(r,r_*)}
\end{equation}
for $\theta\leq \pi/4$.  For $\theta\geq \pi/4$, there should be a plus sign
on the right-hand side.  Integrating Eq.~(\ref{rtheta}) we arrive at the
explicit equation for $\theta=\theta(r)$
\begin{equation}                                                             
\label{rsol}                                                             
\theta=\int_{r}^\infty {d\rho}\,\rho^{-2}\,e^{-\rho^2/4}\,[F(\rho,r_*)]^{-1/2}
\end{equation}
for $\theta\leq \pi/4$.  For $\pi/4<\theta<\pi/2$, the interface is symmetric
with respect to the diagonal, that is, $r(\theta)=r(\frac{\pi}{2}-\theta)$.
The solution in Eq.~(\ref{rsol}) contains the unknown $r_*$, which is the
scaled distance from the origin to the closest point on the interface.  Its
value is obtained by ensuring that $\theta=\pi/4$ when $r=r_*$.  This gives
the criterion
\begin{equation}                                                             
\label{r*}
\int_{r_*}^\infty {dr}\,r^{-2}\,e^{-r^2/4}
[F(r,r_*)]^{-1/2}=\frac{\pi}{4}\,,
\end{equation}
whose numerical solution is $r_*\approx 1.0445$.  Equation (\ref{rsol}), with
$F$ given by (\ref{F}), provides an explicit representation of $\theta(r)$ on
the interface in terms of the (scaled) distance $r\in[r_*,\infty)$ from the
origin.

In the asymptotic regime $r\to\infty$, the form of the interface becomes much
simpler.  {}From Eqs.~(\ref{F}) and (\ref{rsol}) we find
\begin{equation}                                                             
\label{rasymp} 
\theta\to A\,r^{-3}\,e^{-r^2/4} 
\end{equation} 
with $A=2\,[F(\infty,r_*)]^{-1/2}\approx 2.74404$.  Equivalently,
\begin{equation}                                                             
\label{Yasymp} 
Y\to A\,X^{-2}\,\exp\left[-\frac{X^2}{4}\right]\,.  
\end{equation} 
Apart from the numerical factor $A$, this behavior can be established
directly from Eq.~(\ref{Y}) after dropping the subdominant terms in the
asymptotic limit.

While the spatial extent of the continuum interface, defined as the region
with non-zero curvature, is strictly infinite, the presence of a lattice
cutoff implies that the interface will have a finite extent.  The finiteness
of the interface may be quantified by the distance of its leading edge,
$x_{\rm max}$ (or $y_{\rm max})$, from the origin (see Fig.~\ref{def}).  We
may estimate this distance as the value of $x$ for which the TDGL description
first gives $y(x)<a$, where $a$ is the lattice spacing.  Substituting the
criterion $y=a$ into Eq.~(\ref{Yasymp}) we thereby obtain
\begin{eqnarray*}
\frac{a}{\sqrt{Dt}}=\frac{A\, Dt}{x_{\rm max}^2}\,\,e^{-x_{\rm max}^2/4Dt}\,, 
\end{eqnarray*}                                                             
which leads to the asymptotic behavior
\begin{equation}                                                             
\label{xmax} 
x_{\rm max}\approx \sqrt{2Dt\,\ln(Dt/a^2)}.
\end{equation} 
Notice that the value of the lattice spacing is immaterial for the
asymptotic behavior and we therefore set $a=1$ henceforth.

\section{Microscopic Description}
\label{ising} 

\subsection{Basic Characteristics and Generalizations}

\begin{figure}[ht] 
 \vspace*{0.cm}
 \includegraphics*[width=0.18\textwidth]{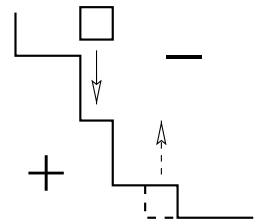}
\caption{The elementary steps of deposition and evaporation that drive 
  the evolution of a staircase.  In this example, deposition (equivalent to
  the spin flip event $-\to +$) can occur at 4 sites while evaporation can
  occur at 3 sites.
\label{process}}
\end{figure}

At the microscopic level, the interface has a staircase shape
(Fig.~\ref{def}).  Zero-temperature Glauber spin-flip dynamics \cite{glauber}
forbids energy raising flips, so that only spins in the corners on the
interface can evolve.  While both energy decreasing and energy conserving
flips are generically allowed, only energy conserving spin flips can occur in
the wedge geometry.  We define the rate for these events to be one without
loss of generality.  The construction of the system ensures that the total
number of possible flips of minority spins always exceeds the total number of
possible flips of majority spins by one (see Fig.~\ref{process}).  Hence the
total number $S_t$ of spins in the first quadrant that join the majority
phase is a random variable that undergoes a random walk on the half-lattice
$S_t\in {\bf Z}_+$, with a constant positive bias that equals one.  Thus the
expected number of spins that have flipped at time $t$ is $\langle
S_t\rangle=t$.

Algorithmically, Glauber kinetics involves randomly picking a spin on the
corners of the staircase and allowing this spin to flip freely.
Equivalently, we can view the staircase evolution as a deposition/evaporation
process in which deposition can occur at sites $x$ where $y(x)<y(x-1)$ (with
$y(-1)$ defined to be infinite), while evaporation can occur at sites where
$y(x)>y(x+1)$ (Fig.~\ref{process}).  Here deposition is equivalent to the
spin flip process $-\to +$ and {\it vice versa} for evaporation.

According to zero-temperature Glauber kinetics, deposition and evaporation
events must occur at the same rate for all eligible sites.  Owing to the
above-mentioned fact that there is always exactly one more site
available for deposition than for evaporation, this ``unbiased'' evolution
rule leads to a steadily growing interface in which the average number of
particles in the deposit grows as $t$.  This particulate description for the
interface naturally suggests the generalization to different deposition and
evaporation rates.  For reasons that will soon become evident, we consider
the following three rules:

\begin{itemize}
  
\item[(i)] Equal deposition and evaporation rates.  This is just evolution of the
  interface by Glauber kinetics at zero temperature and zero magnetic field.
  
\item[(ii)] Evaporation rate greater than deposition rate.  With this rate bias,
  the interface reaches an equilibrium state.
  
\item[(iii)] Irreversible deposition with no evaporation events.  
  
\end{itemize}

Physically, rule (iii) is equivalent to interface evolution by
zero-temperature Glauber kinetics in the presence of a magnetic field that
favors the majority spin; the magnitude of the field is irrelevant (at zero
temperature) as long as it is smaller than a threshold value to ensure the
stability of flat interfaces.

\subsection{Relation to Partitions}

\begin{figure}[ht] 
 \vspace*{0.cm}
 \includegraphics*[width=0.18\textwidth]{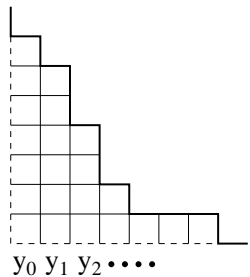}
\caption{The Young diagram that is based on the interface profile of 
  Fig.~\ref{def}.  This diagram corresponds to a partition of the integer 22
  into the set $\{7,6,4,2,1,1,1\}$.
\label{young}}
\end{figure}

The staircase can also be viewed as a geometric representation of the
partition of the integer number $S_t$.  This partition is simply the set of
$S_t$ boxes in the first quadrant that are arranged in non-increasing order.
Such an object is also called a Young diagram \cite{Fulton}.  For example,
the interface of Fig.~\ref{young} corresponds to the partitioning of the
integer 22 into the set $\{7,6,4,2,1,1,1\}$.

Much is known about partitions \cite{Andrews}; for example, the number of
partitions $p(N)$ of the integer $N$ has the asymptotic behavior $p(N)\sim
N^{-1}\,\exp(2\pi\sqrt{N/6})$.  The asymptotic behavior of Young diagrams is
also qualitatively simple: After a suitable rescaling, a typical Young
diagram converges to a {\em limiting shape} (see \cite{Ver} for precise
statements and earlier references).  To compute this shape one must know the
weights of all possible partitions.  For the case where each partition occurs
with equal weight $1/p(N)$ (the uniform measure), the corresponding limiting
shape is known \cite{Ver}, and will be quoted in the next subsection.

For the Ising interface (rule (i) in the above list), we do not know the
weights of each interface configuration \cite{fluct}.  We will see that the
shape of the Ising interface is slightly different from that of Young
diagrams, implying that the weights are not uniform.  On the other hand, if
the spin flip event $+\to -$ is favored over $-\to +$ (rule (ii)), then the
interface approaches an equilibrium state.  Because the state space is
sampled more extensively in the equilibrium system, this suggests that the
weights for each interface configuration should be more uniform than in the
case of rule (i), as borne out in our simulation results below.

\subsection{Limiting Shape}

We now present a heuristic derivation for the limiting shape that corresponds
to the uniform measure.  We follow an argument by Shlosman \cite{shl}; see
also \cite{rd} for a similar approach.  The assumption that the measure is
uniform implies that we can disregard the underlying dynamics and simply
count the number of possible staircases.  (A similar calculation for the
triangular lattice is presented in the Appendix.)

On the square lattice, the interface is a staircase in ${\bf Z}^2$ with each
step going either to the right or downward (Fig.~\ref{def}).  Let ${\bf
  1}=(x_1,y_1)$ and ${\bf 2}=(x_2,y_2)$ be two points in ${\bf Z}^2$ that can
be connected by a staircase; that is, $0\leq x_1\leq x_2$ and $y_1\geq
y_2\geq 0$.  The number of staircases from ${\bf 1}$ to ${\bf 2}$ is
\begin{equation}                                                             
\label{N12-def} 
N({\bf 1},{\bf 2})={x_2-x_1+y_1-y_2\choose x_2-x_1}\,.      
\end{equation} 
For compactness, we set $a=x_2-x_1$ and $b=y_1-y_2$.  If the points ${\bf 1}$
and ${\bf 2}$ are distant, {\it i.e.}, $a\gg 1$ and $b\gg 1$, the Stirling
formula gives
\begin{eqnarray*}               
\ln {a+b\choose a}\longrightarrow -a\ln\frac{a}{a+b}-b\ln\frac{b}{a+b}\,.
\end{eqnarray*}                                                             
Therefore
\begin{equation}
\label{ab}                
\ln N({\bf 1},{\bf 2})=\ln {a+b\choose a}
\longrightarrow \sqrt{a^2+b^2}\,\,\Phi({\bf n})\,,
\end{equation}                                                             
where ${\bf n}=(n_1,n_2)=(b,a)/\sqrt{a^2+b^2}$ is the unit vector orthogonal
to ${\bf 1}-{\bf 2}$ and 
\begin{equation}                                                             
\label{fn12} 
\Phi({\bf n})=-n_1\,\ln\frac{n_1}{n_1+n_2}-n_2\,\ln\frac{n_2}{n_1+n_2}\,. 
\end{equation} 

Suppose now that points ${\bf 1}$ and ${\bf 2}$ are far enough apart to
ensure the applicability of Eq.~(\ref{ab}) yet close enough to guarantee
that the interface is locally almost flat.  Under these conditions, we have
\begin{equation}                                                             
\label{n12} 
(n_1,n_2)=\left(-\frac{y_x}{\sqrt{1+y_x^2}}\,,\frac{1}{\sqrt{1+y_x^2}}\right).
\end{equation} 

Generally, consider an interface that goes through the points ${\bf
  1},\ldots, {\bf k}$, with adjacent points satisfying the above
requirements.  The total number of these staircases is then the product of
the factors $N({\bf j},{\bf j+1})$.  The logarithm of the number of
staircases is therefore the sum of these factors (asymptotically an
integral).  Using Eqs.~(\ref{ab}) and (\ref{n12}) we thereby find that the
logarithm of the total number of staircases approaches to
\begin{eqnarray}                                                             
\label{N} 
G[y]&=&\int_0^\infty dx\,\sqrt{1+y_x^2}\,\,\,
\Phi\left(-\frac{y_x}{\sqrt{1+y_x^2}}\,,
\frac{1}{\sqrt{1+y_x^2}}\right)\nonumber\\
&=&\int_0^\infty dx\,
\left[y_x\,\ln\frac{-y_x}{1-y_x}-\ln\frac{1}{1-y_x}\right].
\end{eqnarray} 
Since $x, y\propto\sqrt{t}$ on the interface, the number of staircases near
the typical interface scales as $e^{\sqrt{t}\,G[Y]}$.  Because $G[Y]$ is of
the order of one, the number of staircases rapidly grows with $t$ and the
dominant contribution arises from the staircases close to the curve
$y=y(x,t)$ that maximizes the functional of Eq.~(\ref{N}).  Thus we need to
determine only the optimal curve to find the asymptotic interface shape.

\begin{figure}[ht] 
 \vspace*{0.cm}
 \includegraphics*[width=0.38\textwidth]{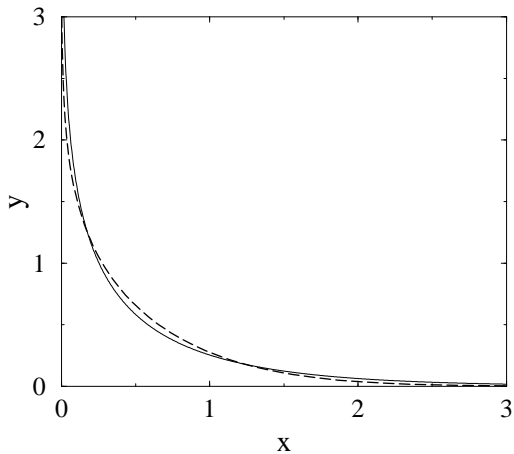}
\caption{Comparison of the interfaces from the TDGL equation
  (Eq.~(\ref{rsol})) (dashed) and that from Eq.~(\ref{main}) (solid).  Both
  curves are normalized to unit enclosed area.
\label{cont-vs-disc}}
\end{figure}

In maximizing (\ref{N}) we must only use curves that bound an area equal to
$t$, {\it i.e.}, $\int_0^\infty dx\,y(x,t)=t$.  With this isoperimetric
constraint, the proper functional to maximize is $G_\lambda[y] \equiv
G[y]-\lambda\int_0^\infty dx\,y$, where $\lambda$ is a Lagrange multiplier.
Re-writing $G_\lambda[y]= \int_0^\infty dx\,L(y,y_x)$, with the Lagrangian
\begin{equation}
\label{lagrange}                
L(y,y_x)=y_x\,\ln\frac{-y_x}{1-y_x}-\ln\frac{1}{1-y_x}
-\lambda y,
\end{equation}                                                             
and applying the Euler-Lagrange formalism, gives the extremum condition
$\frac{d}{dx}\,\ln\frac{-y_x}{1-y_x}=-\lambda$.  Integrating this equation
subject to $y_x(0)=-\infty$ yields
\begin{equation}
\label{Yx}                
\frac{-y_x}{1-y_x}=e^{-\lambda x}\,.
\end{equation}                                                             
Integrating (\ref{Yx}) subject to $y(\infty)=0$ gives the remarkably simple
form \cite{shl} for the shape of the optimal staircase
(fig.~\ref{cont-vs-disc})
\begin{equation}                                                             
\label{main} 
e^{-\lambda x}+e^{-\lambda y}=1, \quad {\rm with}\qquad \lambda=\frac{\pi}{\sqrt{6t}}\,,
\end{equation} 
where $\lambda$ is determined from the area constraint.

This limiting shape is quantitatively close to that obtained from the
coarse-grained TDGL approach, except near the tail region, where the TDGL
equation predicts a Gaussian tail (\ref{Yasymp}) while (\ref{main}) gives an
exponential tail
\begin{equation}                                                             
\label{Yasympexact} 
Y\to \frac{\sqrt{6}}{\pi}\,\exp\left[-\frac{\pi}{\sqrt{6}}\,X\right], 
\end{equation} 
as $X\to\infty$, where $X=x/\sqrt{t}$ and $Y=y/\sqrt{t}$ are the scaled
coordinates.  Correspondingly, the location of the leading edge is
\begin{equation}                                                             
\label{xmaxexact} 
x_{\rm max}\to C\sqrt{t}\,\,\ln t
\end{equation} 
with $C=\sqrt{6}/2\pi=0.3898\ldots$.  Notice that the leading edge moves
slightly faster than the $\sqrt{t\,\ln t}$ law predicted by the TDGL
approach.

\subsection{Simulation Results}

We simulated Ising interfaces that are grown by the three different rules
defined in Sec.~III.A.  By rescaling each of these interfaces to have unit
area, our numerical results exhibit data collapse after a short-time
transient, with each rule giving a slightly different, although
quantitatively similar, universal curve.  Interestingly, only in the
evaporation-dominated case (rule (ii)), does the interface shape coincide
with Eq.~(\ref{main}).

\begin{figure}[ht] 
 \vspace*{0.cm}
 \includegraphics*[width=0.42\textwidth]{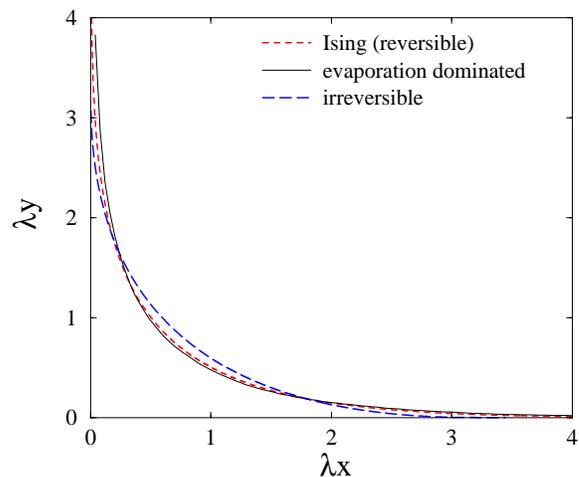}
 \includegraphics*[width=0.42\textwidth]{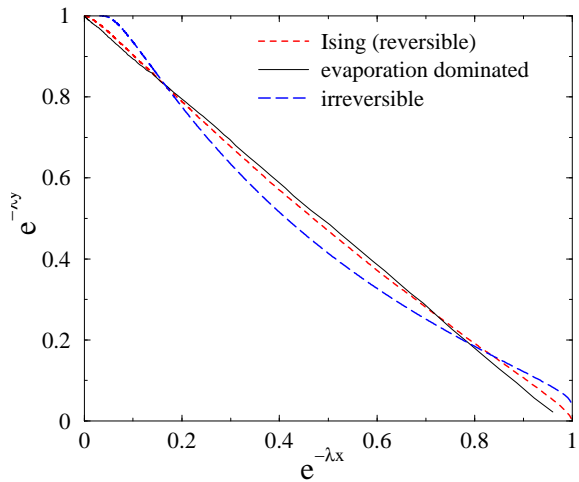}
\caption{The interface in the wedge geometry for: (i) equal deposition
  and evaporation rates --- equivalently, the Ising-Glauber interface (based
  on 100 realizations of $434033 1\approx 1.5^{32}$ particles), (ii)
  evaporation rate exceeding the deposition rate (based on $500$ realizations
  and the ratio of evaporation to deposition rate equal to $0.51/0.49\approx
  1.041$, corresponding to an average deposit of 1012 particles), and (iii)
  irreversible deposition (based on $10^4$ realizations of $287627=1.5^{31}$
  particles).  The two plots show the interface on a linear (top) and an
  exponential scale (bottom).  The straight line behavior in the latter
  corresponds to Eq.~(\ref{main}).
\label{compare}}
\end{figure}
 
It is also worth noting the following subtlety in our measurements of
interfaces.  In each realization, we record the height $y(x)$ at each value
of $x$, and then we average over many realizations to obtain the average
interface $\{\langle y(x)\rangle\}$.  This procedure is manifestly not
symmetrical about the $45^\circ$ diagonal.  For example, in our definition of
the average profile there is necessarily a non-zero contribution at $x_{\rm
  max}$, so that the average profile in the $x$-direction extends to $x_{\rm
  max}$.  On the other hand, the average profile in the $y$-direction extends
only to the smaller value $\langle y(x=1)\rangle$.  The asymmetry caused by
this averaging is small, except near the extremes of the interface (if one
looks closely at Fig.~\ref{compare}).

\section{Related Geometries}

The next level of complexity is to consider an initial interface with two
corners.  We specifically study two cases: (a) a single large step that
smooths out to an error function profile, and (b) a semi-infinite rectangular
``finger'' that evolves to a constantly receding steady shape.

\subsection{Single step}

\begin{figure}[ht] 
 \vspace*{0.cm}
 \includegraphics*[width=0.32\textwidth]{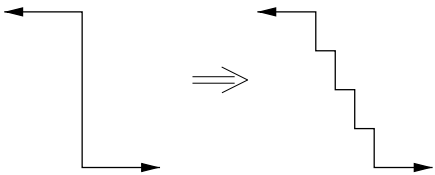}
\caption{Schematic evolution of a single step according to Ising-Glauber 
  dynamics.
\label{one-step}}
\end{figure}

We form a single step interface by two horizontal half-lines, $x\leq 0, y=2h$
and $x\geq 0, y=0$, and the vertical interval $x=0, 0\leq y\leq 2h$
(Fig.~\ref{one-step}).  If the height $2h$ of the step is small, the problem
is best analyzed by random walk techniques. For example, when $2h=2$, the
kink in the interface is equivalent to a single particle that undergoes a
discrete one-dimensional random walk.  Similarly when $2h=4$ the system
consists of two random walks, located say at $x_L$ and $x_R$, but subject to
the constraint that the two particles cannot interchanges their positions.

More interesting behaviors occur for $h\gg 1$, where three regimes arise.
When $1\ll t\ll h^2$, the two corners do not `feel' each other and the
problem reduces to that of two non-interacting quadrants.  We may obtain a
better upper bound for this non-interacting regime by using the growth law
for the position of the leading edge (Eq.~(\ref{xmax})).  Since the
unperturbed initial vertical interval starts at $y\approx C\sqrt{t}\,\ln t$
and ends at $y\approx 2h-C\sqrt{t}\,\ln t$, the two corners remain
independent as long as $\sqrt{t}\,\ln t\ll h$, or $t\ll h^2/\ln h$.

When $h\sim \sqrt{t}$, the corners interact.  In this second regime, we can
still determine the interface shape by employing the maximization procedure
of section \ref{ising}.  To find $y=y(x,t)$ for $x>0$, we should maximize the
functional $G_\lambda[y]=\int_0^\infty dx\,L(y,y_x)$, with the Lagrangian
given by (\ref{lagrange}), but now with the boundary conditions $y(0)=h$ (by
symmetry) and $y(\infty)=0$.  We find
\begin{equation}                                                             
\label{step} 
\left(1-e^{-\lambda h}\right)e^{-\lambda x}+e^{-\lambda y}=1.
\end{equation} 
The area bounded by this curve is 
\begin{equation}                                                             
\label{steparea} 
\int_0^\infty dx\,y=\lambda^{-2}\,{\rm Li}_2\left(1-e^{-\lambda h}\right),
\end{equation} 
where ${\rm Li}_2(z)=\sum_{n\geq 1} z^n/n^2$ is the dilogarithm function.
Equating the area to $t$ we obtain
\begin{equation}                                                             
\label{relation} 
\Lambda^{-2}\,{\rm Li}_2\left(1-e^{-\Lambda H}\right)=1,
\end{equation} 
where $\lambda=\Lambda/\sqrt{t}$ and $H=h/\sqrt{t}$.  In the limit $H\to
\infty$, we must recover the non-interacting regime; indeed, the limiting
shape (\ref{step}) reduces to (\ref{main}).

In the third regime $h\ll\sqrt{t}$, the most appropriate description of the
interface is in terms of $2h$ random walkers.  Adjacent walkers are separated
by a large distance of the order of $\sqrt{t}/h$ and therefore are
effectively non-interacting.  We may compute the density of random walkers by
solving the diffusion equation and the resulting limiting shape is given by
the error function.  This same prediction follows from the TDGL approach, as
the factor $y_x^2$ can be neglected in the long-time limit (see
Eq.~(\ref{y})).  Notice, however, that it is not possible to recover this
limiting shape by taking the $H\to 0$ limit in Eq.~(\ref{step}).  The reason
for this non-analyticity is the large discrepancy between the horizontal and
vertical scales when $\sqrt{t}\gg h$.

\subsection{Rectangular finger}

For the finger geometry, the minority phase initially occupies the
semi-infinite region $y>0$ and $|x|<L$.  The interesting regime is again
$t\gg L^2$, where the two corners of the initial finger interact and the
finger relaxes to a limiting shape that eventually recedes at constant
velocity.  In a reference frame moving with the finger, the interface $y(x)$
is thus stationary.

\begin{figure}[ht] 
 \vspace*{0.cm}
 \includegraphics*[width=0.32\textwidth]{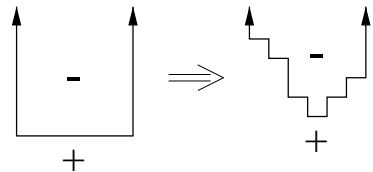}
\caption{Schematic evolution of a rectangular finger according to Ising-Glauber 
  dynamics.  On the right-hand side, the flip of the lowest minority spin is
  an irreversible process that causes the minimum height of the finger to
  advance by one.
\label{finger-pic}}
\end{figure}

A new feature of the semi-infinite finger compared to the wedge geometry is
the possibility for energy-lowering spin-flip events to occur.  For example,
if the tip of the finger contains a single spin, then when this spin flips,
the fingertip irreversibly advances by one unit.  Another new feature is that
the finger can shed disconnected pieces whenever the tip of the finger has a
width equal to one and length greater than two.  Finally, the possibility of
energy-lowering moves also means that the evolution of a finger is
irreversible; there is no possibility of the system returning to its initial
state once an energy-lowering move has occurred (Fig.~\ref{finger-pic}).

Within the TDGL approach, $y(x)$ now satisfies the equation
$Dy_{xx}=v(1+y_x^2)$ \cite{Mu,PMVC}.  Integrating, and imposing the boundary
condition $y\to\infty$ when $|x|\to L$, we obtain, for the finger shape
\begin{equation}
\label{finger}
y=-\frac{2L}{\pi}\,\ln\left[\cos\left(\frac{\pi x}{2L}\right)\right],
\end{equation}
a result that was first apparently obtained by Mullins \cite{Mu}.
In this steady state, the finger recedes at a constant velocity that is given
by $v=Dy''(0)=\pi D/2L$.  

Again, we test the applicability of the TDGL approach by comparing with
numerical simulations.  In our simulations of the finger, both energy
conserving and energy lowering moves can occur, with rates 1 and 2,
respectively.  The rate at which the finger recedes is controlled by the fact
that there is almost always an excess of two sites where the spin flip event
$-\to +$ can occur (recession step) compared to $+\to -$.  The only exception
is the case where the fingertip width equals one; here the excess of
potential recession steps over advancement steps also equals one.  As a
result of this nearly constant bias, the finger recedes at a constant rate
with steady-state velocity equal to $1/L$, up to exponentially small
corrections.

\begin{figure}[ht] 
 \vspace*{0.cm}
 \includegraphics*[width=0.42\textwidth]{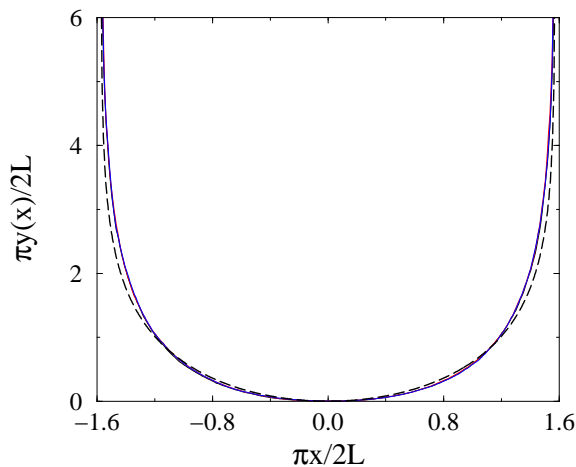}
\caption{Comparison of the finger shape predicted by the TDGL approach given
  in Eq.~\ref{finger} (dashed line) with simulation results for 100
  realizations of width $2L=400$ at times $t=2\times 10^5$, $6\times 10^5$,
  and $10^6$ steps (solid lines).  The data for these three times essentially
  coincide.
\label{finger-compare}}
\end{figure}

While the shape of the finger is quantitatively close to the TDGL prediction,
the discrepancy between the continuum theory and the simulations persists
even as $L\to\infty$.  This dichotomy is parallel to that observed in the
wedge geometry.

We have not investigated the evolution of the interface subject to rules (ii)
and (iii) in detail, so we just mention qualitative results.  In both cases,
the finger recedes at a constant velocity.  For the system in a magnetic
field (rule (iii)), $v(L)$ is a decreasing function of $L$ that saturates to
a non-zero limit $v_\infty=\lim_{L\to\infty}v(L)>0$, in contrast to the case
finger evolution in zero magnetic field.  In the evaporation-dominated case,
we estimate the velocity as the probability for the finger to reach the state
where the fingertip has a width equal to one.  In this case, an irreversible
energy-lowering move can occur, so that the finger recedes by one step.  If
$1+\epsilon$ is the ratio of the evaporation to deposition rate, then
$v(L)\sim e^{-\epsilon L}$.

\section{Fluctuations}

\begin{figure}[ht] 
 \vspace*{0.0cm}
 \includegraphics*[width=0.38\textwidth]{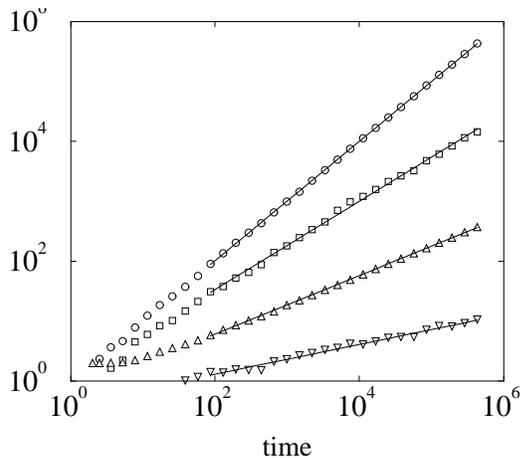}
\caption{Time-dependent properties of the Ising interface with Glauber
  kinetics for the wedge initial condition based on 100 realizations.  Shown
  are: the average interface area ($\circ$), the variance in this area
  ($\square$), the mean value of the closest distance from the origin to the
  interface ($\Delta$), and the variance in this distance ($\nabla$).  The
  thin solid lines are the best fits to the data at long time and have
  respective slopes 0.997, 0.737, 0.490, and 0.248.
\label{time-dep}}
\end{figure}

In addition to the mean interface shape, we also study fluctuations of the
interface.  Perhaps the simplest such quantity is the fluctuation in the area
$S_t$ bounded by the interface at time $t$.  We estimate the probability
distribution of the area by the following simple argument.  In the long-time
limit, there are $N_+\sim t^{1/2}$ spins along the interface that can flip
and join the majority phase and $N_-=N_+-1$ interface spins that can flip and
join the minority.  Thus the evolution of $S_t$ is driven by a deterministic
contribution of rate one and a random contribution whose rate is of the order
of $t^{1/4}$.  This suggests that the evolution of $S_t$ is governed by the
Langevin equation:
\begin{equation}
\label{Langevin}                
\frac{dS_t}{dt}=1+t^{1/4}\,\xi(t),
\end{equation}
with $\xi(t)$ a random noise term that satisfies $\langle \xi(t)\rangle=0$
and $\langle \xi(t)\,\xi(t')\rangle=\delta(t-t')$.  {}From this equation we
immediately find that $\langle S_t\rangle=t$, while the fluctuation in $S_t$
is given by
\begin{eqnarray*}               
\langle (S_t-t)^2\rangle
&=&\int_0^t dt_1\int_0^t dt_2\, (t_1t_2)^{1/4}
\langle \xi(t_1)\,\xi(t_2)\rangle\\
&=&\int_0^t dt_1\,\,t_1^{1/2}\propto t^{3/2}\,.
\end{eqnarray*}
Thus $S_t-t$ is a Gaussian random variable, and the probability distribution
of the area $P_n(t)\equiv {\rm Prob}[S_t=n]$ is
\begin{equation}                                                             
\label{Pnt}                
P_n(t)\propto \exp\left[-\frac{(n-t)^2}{t^{3/2}}\right]\,.
\end{equation} 
The variance is proportional to $t^{3/4}$, in excellent agreement with our
simulation results (Fig.~\ref{time-dep}).  As a corollary to this latter
result, the probability $P_0(t)$ for the interface to return to its original
state decays as $P_0(t)\propto e^{-\sqrt{t}}$.  Because $\int P_0(t')\, dt'$
is finite, this means that the probability for the interface to eventually
return to its original state is less than one \cite{fpp}.

We next consider fluctuations in the position of the interface by studying
the location of the intersection of the interface with the diagonal $x=y$.
The intersection between the diagonal and the (deterministic) limiting shape
is $x=y=C_1\sqrt{t}$ with $C_1=(\sqrt{6}\,\ln 2)/\pi$.  In each realization,
however, $x$ is a random variable.  Following the same argument as that
applied for the total interface area, we anticipate that the variance in the
position of the intersection point exhibits Gaussian fluctuations.  This
gives
\begin{equation}                                                             
\label{x-fluct}                
\left|x-C_1\sqrt{t}\right|\propto t^{1/4}\,,
\end{equation} 
again in excellent agreement with simulations (Fig.~\ref{time-dep}).

\section{Discussion}

The dynamics of a single unbounded interface between ordered phases in the
two-dimensional zero-temperature Ising-Glauber model has surprisingly rich
properties and exciting connections with diverse topics in mathematics and
physics.  We discussed the correspondence to the limiting shapes of
partitions \cite{Ver,O}; two other connections are to exclusion processes
\cite{excl}, and potentially to random matrices \cite{mehta}.  In particular,
we presented evidence that in the case where the interface achieves an
equilibrium state, the resulting interface coincides with the limiting shape
in the partitioning problem.

\begin{figure}[ht] 
 \vspace*{0.cm}
 \includegraphics*[width=0.32\textwidth]{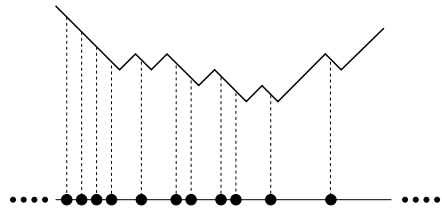}
\caption{The interface configuration of Fig.~\ref{young} rotated by
  $45^\circ$ and the corresponding particle configuration.
\label{csp}}
\end{figure}

The connection between the interface in the wedge geometry and exclusion
processes arises from a simple construction in which one associates
a particle with each vertical portion of the interface and a hole with each
horizontal portion (Fig.~\ref{csp}).  Thus the wedge geometry corresponds to
an initial state in the particle system that consists of a semi-infinite line
of particles in the region $(-\infty,0)$ and empty space for $(0,\infty)$.
Basic features of the interface shape can therefore be translated to
corresponding properties of the particle density profile.

A possible connection to random matrices is best appreciated for the
situation where the Ising interface grows irreversibly (our rule (i)).  In
this case, the weight of any interface configuration is the number of ways to
grow the final state from the empty diagram $\emptyset$ by adding squares one
at a time, such that a partition is maintained at each step.  For each
partition $\pi=(y_0\geq y_1\geq \ldots\geq 0)$ of size $|\pi|=\sum y_j=N$,
the number of distinct growth histories is usually denoted $\dim \pi$
\cite{dim} and irreversible interface growth corresponds to finding the
limiting shape with the probability measure
\begin{equation}                                                             
\label{measure} 
{\rm Prob}(\pi)=\frac{\dim \pi}{Z_N}\,, 
\end{equation} 
where $Z_N=\sum_{|\mu|=N} \dim \mu$ \cite{ZN}.  For the closely related
Plancherel measure, ${\rm Prob}(\pi)\propto (\dim \pi)^2$, the limiting shape
is also known \cite{VK,LS}.  As explained in Ref.~\cite{O}, the simplicity of
the Plancherel measure stems from a hidden connection to unitary Gaussian
random matrices; similarly, the measure (\ref{measure}) appears to be related
to orthogonal Gaussian random matrices.

Another open question is the limiting shape of an initially large rectangle
of down spins in an infinite sea of up spins.  As mentioned in the
introduction, although the time scale for this object to disappear is known
\cite{lif}, the {\it shape} of this object is not.  One might expect that an
initial square would evolve to a circular shape.  It has indeed been proved
that for curvature-driven growth every smooth closed curve in the plane
asymptotically approaches a (shrinking) circular shape \cite{GH86,G87}.  The
situation in three dimensions is much richer because a surface with both
concave and convex portions can undergo fission by curvature-driven growth.
The analogous result to Grayson's theorem is that any convex domain will
ultimately approach a shrinking sphere \cite{H84}.  On the other hand, for
the Ising-Glauber model, nothing has been established rigorously.  Our
analytic and numerical results showed that lattice anisotropy effects persist
in the interface dynamics.  Thus we expect that a shrinking cluster in the
long-time limit will not be isotropic.

Moving to three dimensions, it should be worthwhile to investigate the shape
of the zero-temperature Ising-Glauber interface on the cubic lattice when the
spins in the positive octant $x\geq 0, y\geq 0, z\geq 0$ have a different
sign than all other spins.  We anticipate that the results will be similar to
the limiting shape of the so-called plane partitions.  In this latter
problem, analytical results recently obtained in Refs.~\cite{CK,OR}
correspond to the case of the uniform measure.  It would be interesting to
study the possible correspondence between these plane partitions and the
Ising interface.

Finally, we cannot resist mentioning that the finger shape (\ref{finger}) is
mathematically identical to the Saffman-Taylor finger in the Hele-Shaw cell
\cite{ST}.

\medskip\noindent We are grateful to K.~Kornev and R.~Rajesh for helpful
remarks and especially to B. Meerson for helpful comments on the manuscript
and for bringing Ref.~\cite{Mu} to our attention.  PLK and SR thank NSF grant
DMR0227670 for financial support of this research.  The research of JT was
supported by a travel grant from ENS de Cachan.

\appendix
\section{Triangular lattice}

We study the triangular lattice \cite{hex} because, in contrast to the square
lattice, it is macroscopically isotropic, and hence one might anticipate that
interface dynamics should be described by the intrinsically isotropic TDGL
equation.  We will see, however, that this is {\em not\/} the case.

\begin{figure}[ht] 
  \vspace*{0.cm} \includegraphics*[width=0.34\textwidth]{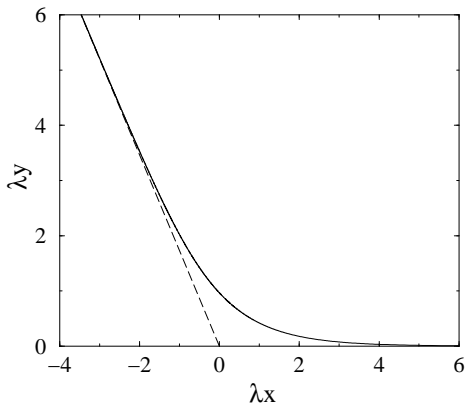}
\caption{The interface for the triangular lattice predicted by Eq.~(\ref{main3})
  for the $2\pi/3$ wedge (dashed line).
\label{triangular}}
\end{figure}

There are two natural possibilities for a wedge geometry on the triangular
lattice: opening angle (a) $\pi/3$ and (b) $2\pi/3$.  The latter case is more
amenable to analysis, since energy decreasing spin flips can never occur and
the system has the same reversibility as the $\pi/2$ wedge on the square
lattice.

Our derivation of the interface shape follows the steps given for the $\pi/2$
wedge on the square lattice.  The only new feature is that
$a=(x_2-x_1)-(y_1-y_2)/\sqrt{3}$ and $b=2(y_1-y_2)/\sqrt{3}$ should be
employed in Eq.~(\ref{ab}).  The variational problem on the triangular
lattice involves maximizing the functional $G_\lambda[y]=\int_0^\infty
dx\,L(y,y_x)$ with the associated Lagrangian
\begin{eqnarray*}               
L(y,y_x)&=&\frac{2}{\sqrt{3}}\,y_x\,\ln\frac{-2y_x}{\sqrt{3}-y_x}\\
&-&\left(1+\frac{1}{\sqrt{3}}\,y_x\right)\ln\frac{\sqrt{3}+y_x}{\sqrt{3}-y_x}
-\lambda y.
\end{eqnarray*}                                          

The Euler-Lagrange equation reduces to 
\begin{equation}
\label{Lagr}                             
\frac{d}{dx}\,\left[\ln\frac{y_x^2}{3-y_x^2}\right]=-\sqrt{3}\,\lambda.
\end{equation}                                                        
Integrating twice (subject to the boundary conditions $y\to 0$ for
$x\to\infty$ and $y\to -\sqrt{3}\,x$ for $x\to -\infty$) yields
\begin{equation}                                                             
\label{main3} 
x=-\frac{y}{\sqrt{3}}-\frac{2}{\sqrt{3}\,\lambda}\,
\ln\left[1-e^{-\lambda y}\right], 
\quad \lambda=\frac{\pi}{3^{3/4}\sqrt{t}}\,,
\end{equation} 
where $\lambda$ is again determined from the constraint that the area
between the limiting shape and the initial wedge equals $t$.  To simplify
this computation, it is useful to write the area in the form $\int_0^\infty
dy\,(x+3^{-1/2}y)=t$.

The limiting shape of the interface is shown in Fig.~\ref{triangular}.  The
asymptotic tails of the interface are again exponentially small. This agrees
with simulations and contradicts to the Gaussian tails predicted by the TDGL
approach.


\begin{thebibliography}{99}
\bibitem{note} 
  During the evolution, small bubbles can branch off the the domain, so
  strictly speaking there may be more than one interface.
\bibitem{fluct} 
  Furthermore, $S_t$ is not a fixed number but a random variable that
  fluctuates from realization to realization; this apparently has an
  asymptotically negligible effect as the relative fluctuations in $S_t$ decay
  to zero (see Sect.~V).
\bibitem{dim} 
  The irreducible representations of the symmetric group $S_N$
  are indexed by partitions $\pi$ of size $N$, and $\dim \pi$ is the dimension
  of the corresponding representation \cite{Fulton}.
\bibitem{ZN} 
  The normalization factor can be found from the generating function
  $\sum_{N\geq 0}\frac{Z_N}{N!}\,t^N=\exp(t+\frac{t^2}{2})$.
\bibitem{hex} The third possible two-dimensional lattice, the hexagonal, has
  uninteresting zero-temperature dynamics because the coordination number is
  odd.  Thus any patch containing a hexagon never disappears.
\end{thebibliography}
\end{document}